\documentclass[utf8]{frontiersFPHY}  

\usepackage{url,lineno,microtype,subcaption} 
\usepackage[onehalfspacing]{setspace} 
\usepackage{natbib}

\def \aap{A\&A}
\def \apjl{ApJ}

\def \mnras{MNRAS}
\def \apj{ApJ}

\def \aj{AJ}

\bibpunct{(}{)}{;}{a}{}{,}

\def\keyFont{\fontsize{8}{11}\helveticabold }
\def\firstAuthorLast{Zwitter} 
\def\Authors{Toma\v{z} Zwitter\,$^{1,*}$}
\def\Address{$^{1}$Faculty of Mathematics and Physics, University of Ljubljana, Ljubljana, Slovenia}
\def\corrAuthor{Faculty of Mathematics and Physics, University of Ljubljana, Jadranska 19, 1000 Ljubljana, Slovenia}

\def\corrEmail{tomaz.zwitter@fmf.uni-lj.si}

\begin{document}
\onecolumn

\title[Gaia space mission and quasars]{Gaia space mission and quasars} 

\author[\firstAuthorLast ]{\Authors} 
\address{} 
\correspondance{} 

\extraAuth{}

\maketitle

\begin{abstract}

\section{}
Quasars are often considered to be point-like objects. This is largely true and allows for an excellent alignment of the optical positional reference frame of the ongoing ESA mission Gaia with the International Celestial Reference Frame. But presence of optical jets in quasars can cause shifts of the optical photo-centers at levels detectable by Gaia. Similarly, motion of emitting blobs in the jet can be detected as proper motion shifts. Gaia's measurements of spectral energy distribution for around a million distant quasars is useful to determine their redshifts and to assess their variability on timescales from hours to years. Spatial resolution of Gaia allows to build a complete magnitude limited sample of strongly lensed quasars. The mission had  its first  public data release in September 2016 and is scheduled to have the next and much more comprehensive one in April 2018. Here we briefly review the capabilities and current results of the mission. Gaia's unique contributions to the studies of quasars are already being published, a highlight being a discovery of a number of quasars with optical jets.

\tiny
 \keyFont{ \section{Keywords:} 
extragalactic astronomy, active galactic nuclei, quasars, quasar spectral energy distribution, extragalactic radio jets, strong gravitational lensing, astrometry, photometry} 
\end{abstract}

\section{The Gaia mission}

\begin{figure}
\begin{center}
\includegraphics[width=10cm,angle=270]{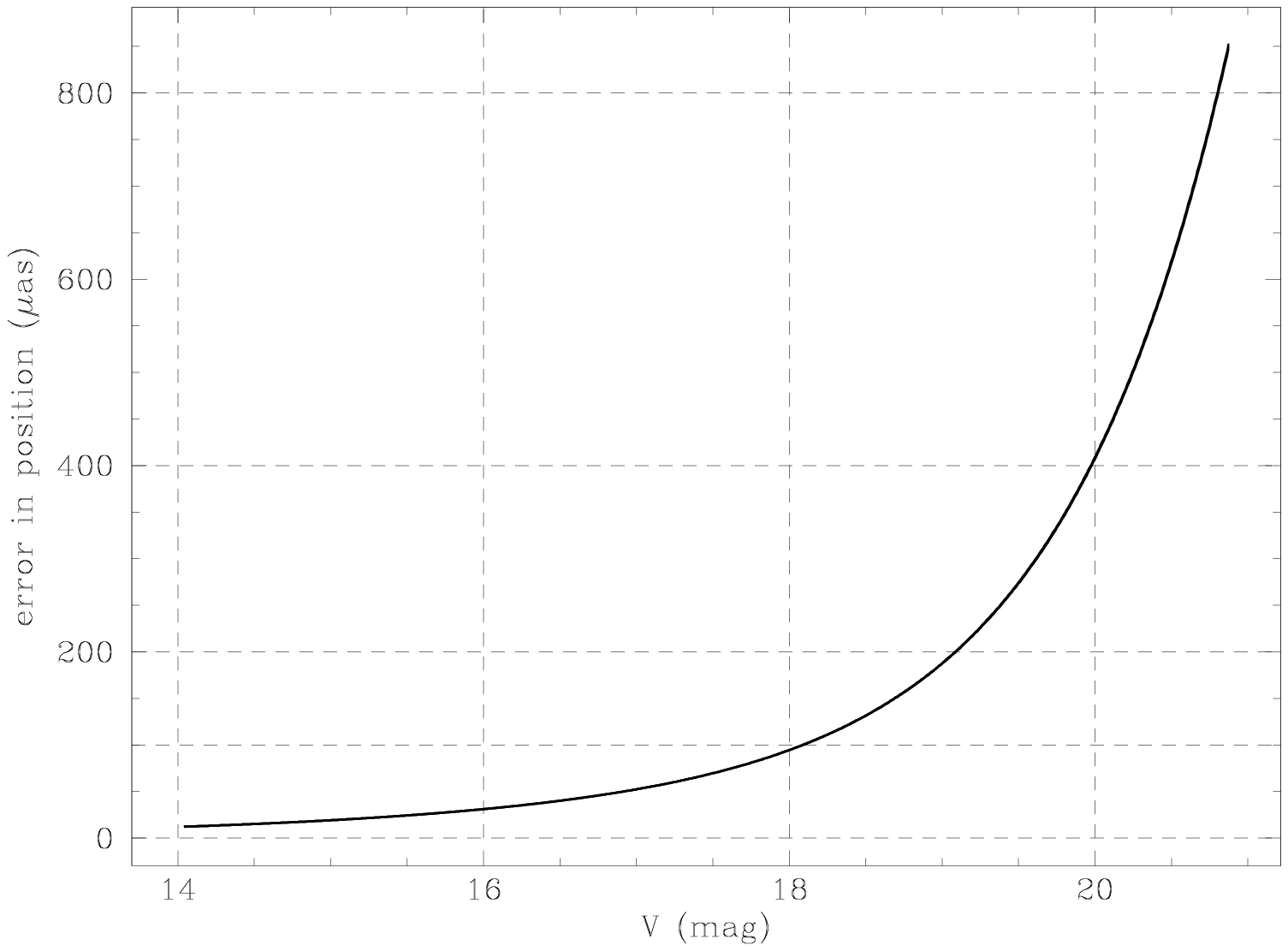}
\end{center}
\caption{Positional error after a 5-year mission as a function of V magnitude. The latter was derived from Gaia's intrinsic $G$ magnitude using the relation $V=1.02 \, G -0.24$, which is appropriate for quasars with $z \leq 4$ \citep{proft15}. Positional error estimate assumes a point-like source and follows equations 4-7 from \citet{prusti16} with $0.54 < V - I < 0.98$.}\label{fig1}
\end{figure}

Gaia, an ongoing mission of ESA (gaia.esa.int), is conducting a massive all sky optical survey of sources with $0.0 \leq V \leq 20.9$ \citep[for current review of mission properties and its status see][]{prusti16}. \citet{brown16}   published an initial sky chart with 1,142,679,969 sources, which already makes Gaia the largest all-sky survey of celestial objects to date. Gaia is primarily an astrometric mission, positional accuracy at the end of the mission is expected to be between 5 and $400~\mu$as (Fig.~\ref{fig1}), while the spatial resolution is $\sim 0.1$~arc-second (as). This impressive accuracy of position of optical photo-center is mostly photon-noise limited and in many cases supersedes the accuracy of VLBI positions for radio emitting sources. Gaia is continuously scanning the skies. In 5 years of official mission lifetime it will observe each source in the sky transiting its focal plane from 50 to over 200 times, median is 72. Sampling is uneven, with the shortest separations corresponding to two observations within the 6-hour spin period of the satellite. Gaia  includes three main instruments: astrometry is done in white light (G band, between 330 and 1050~nm), photometry is collected via weakly dispersed spectra in the blue (BP, blue photometer, 330-680~nm) and red (RP, red photometer, 640-1050~nm) bands, each sampled in wavelength by 45 pixels. 76\%\ of energy of a sharp line from a point source is spread in the wavelength direction over $\sim 1.6$ pixels (BP) and $\sim 3.8$ pixels (RP). Finally there is a spectroscopic instrument on-board which is collecting spectra in the 845-872~nm range for objects brighter than $V \sim 15.5$ and at a resolving power of 10,500. The main goal of spectroscopic instrument is to determine radial velocities, but for the very bright stars ($V \leq 12$) it will also measure abundances of chemical elements with lines in wavelength range of the spectrograph.  Magnitude and wavelength limitations of Gaia's spectroscopic instrument are being complemented by ground-based optical stellar spectroscopic surveys which are aimed to obtain stellar parameters and accurate radial velocities, as well as abundances of individual elements for stars which are not within the reach of Gaia. They can however observe only a small fraction of such sources. The ongoing surveys include RAVE \citep[www.rave-survey.org]{kunder17}, LAMOST \citep[www.lamost.org]{cui12}, Gaia-ESO \citep[www.gaia-eso.org]{gilmore12}, and Galah (www.galah-survey.org, \citealt{deSilva15}, see also \citealt{martell17}). 

\section{Gaia and quasars}

\begin{figure}
\begin{center}
\includegraphics[width=10cm,angle=270]{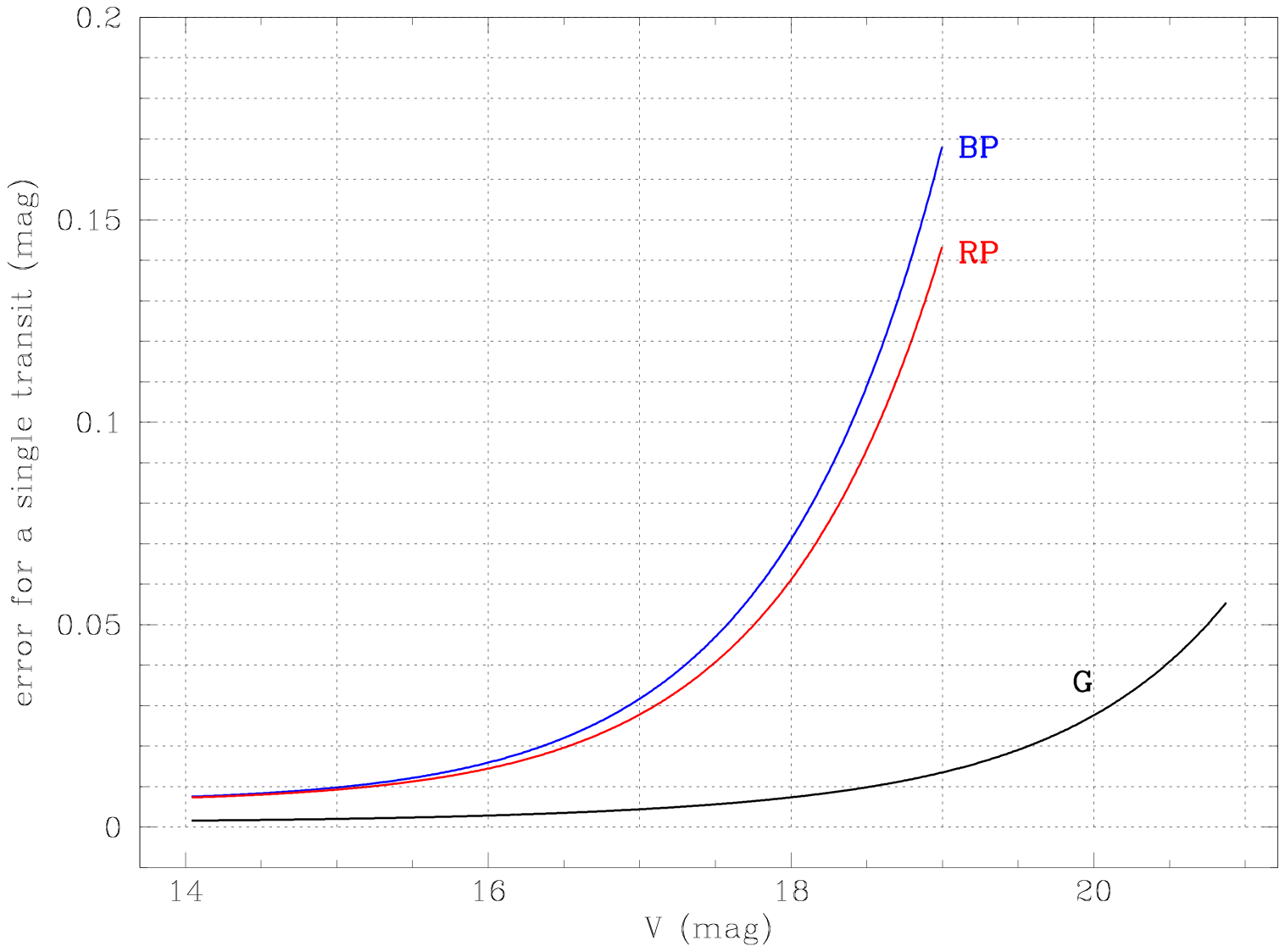}
\end{center}
\caption{Errors in integrated G (white light), RP (red photometer), and BP (blue photometer) magnitudes per transit, as a function of V magnitude, estimated from G magnitude as explained in Fig.\ \ref{fig1}. Errors were estimated from equations 9-13 in \citet{prusti16} using $N_{obs}=1$. Combining blocks of $N_{obs}$ transits obtained on timescales of days or weeks reduces the errors by $\sqrt{N_{obs}}$.}\label{fig2}
\end{figure}

A vast majority of quasars are fainter than $V \sim 15.5$ so they are not observed by the spectroscopic instrument aboard Gaia, but the mission's astrometry and spectrophotometric BP and RP measurements are extremely relevant for quasar studies. Fig.~\ref{fig1} 
shows that one may expect a positional accuracy of $\sim 70 \,\mu$as at $V\sim 17$ and $\sim 400 \,\mu$as at $V\sim 20$ at the end of the official 5-yr mission lifetime. These figures may actually be too pessimistic, as Gaia performs better than expected on the faint end, and the mission itself may be extended up to 2022, therefore increasing the photon budget and time span of its observations. Fig.~\ref{fig2} shows accuracy of integrated magnitudes over the $G$, BP, and RP bands for each transit. It demonstrates the potential of Gaia's $\sim 72$ observations of each quasar over a $\sim 5$~yr timeframe to assess temporal variability for a complete population of quasars down to $V\sim 20$ in white light ($G$ band). For bright sources ($V < 18$) one can also study colour changes on timescales from hours to years, using their integrated BP and RP magnitudes. 

\begin{figure}[h!]
\begin{center}
\includegraphics[width=12cm,angle=0]{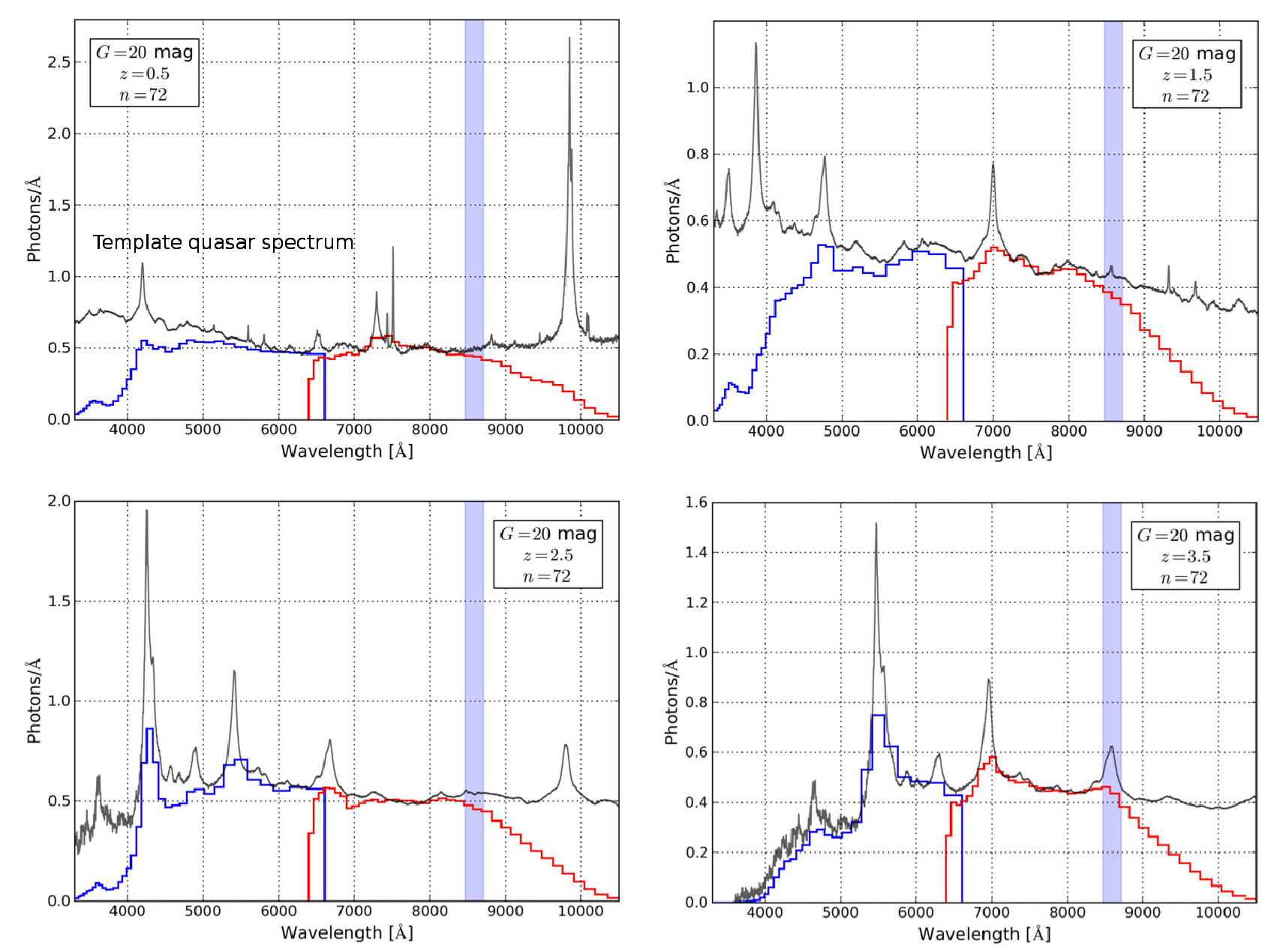}
\end{center}
\caption{Ability of Gaia to determine redshifts of quasars with $G=20$ (corresponding to $V \approx 20.2$) after a median of 72 exposures collected over a 5-year mission. From \citet{proft15}, with permission.}\label{fig3}
\end{figure}

BP and RP instruments do not collect only integrated magnitudes but also spectrophotometric information. Fig.~\ref{fig3} shows a typical quasar spectrum (black tracing), and its BP (blue) and RP (red) spectrophotometry collected by Gaia over a 5-year mission. The panels show the signal for objects with apparent magnitude $V=20$ which have 4 different redshifts between 0.5 and 3.5. Vertical stripe at $\sim8600$~\AA\ is the wavelength range of Gaia's spectroscopic instrument. In our view the figure, which summarises simulations of \citet{proft15}, clearly demonstrates the ability of Gaia to determine photometric redshifts for a complete population of quasars down to $V \sim 20$. For brighter objects even changes in their spectral energy distribution (SED) should be detectable. \citet{proft15} demonstrate that variability of SDSS 154757.71+060626.6, which has Sloan red magnitude $r'=17.5$  and redshift $z=2.03$, is clearly within the reach of Gaia. 

Gaia mission observes any object bright enough to trigger its instruments. So the initial quasar catalogue with 1,248,372 sources, which was prepared before the launch by \citet{andrei14}, is only a lower limit of what we may expect to be observed. Properties of the catalogue are summarised in Table~\ref{table1}. Note the relatively low astrometry precision for most of the quasars. Position is of course more accurately known for 4866 VLBI sources which are part of the catalogue and for additional 38,699 sources with available radio position, although of lower precision. 

\section{Quasars in Gaia's first data release}

Gaia's first data release (Gaia DR1) was published in September 2016 \citep{brown16}. It contains 
\begin{enumerate}
\item
positions, parallaxes (error $\sim 0.3$ milli arc-second, hereafter mas) and proper motions (error $\sim 1$ mas$/$ yr) for 2 million Hipparcos and Tycho stars;
\item
positions (error $\sim 10$~mas) and G magnitudes (error $< 0.03$~mag) for 1.1 billion objects ($V \leq 20.9$);
\item
G-band light curves and characterisation for $\sim 3000$ Cepheids and RR Lyrae stars around the south ecliptic pole.
\end{enumerate}
Quasars observed within the first data release have been discussed by \citet{lindegren16} and by \citet{mignard16}. Main results can be summarised as follows:
Some 135,000 quasars from the list of \citet{andrei14} were included in the astrometric solution, and their positions were determined with an (inflated) standard uncertainty of 1~mas. 
Accurately known positions from VLBI were used to align the Gaia DR1 reference frame with the extragalactic radio frame. 
A special astrometric solution has been computed for selected extragalactic objects with optical counterparts in Gaia DR1, including the ones from the second realisation of the International Celestial Reference Frame (ICRF2). Formal standard error for 2191 such quasars (with $17.6 < V < 20.7$) does not exceed  0.76 mas for 50\%\ and 3.35 mas for 90\%\ of the sources.
Alignment of the Gaia DR1 reference frame with the ICRF2 is better than 0.1~mas at epoch 2015.0. The two frames do not rotate
 to within 0.03~mas$/$yr. There are now 11,444 objects with VLBI positions, i.e. 3.5-times more than in ICRF2 \citep{petrov17}.

\section{Offsets between Gaia and VLBI positions and evidence for optical jets}

\citet{lindegren16} note that for sources with good optical and radio astrometry they found no indication of physical optical vs.\ radio offsets exceeding a few tens of mas, in most cases they are less than 1 mas. This in encouraging, as it will permit to build a very accurate common Gaia$+$radio reference frame in the future. But some objects may require further checks. In particular, 
\citet{makarov17} publish a list of 188 objects (out of 2293 ICRF2 sources), most of them with $z < 0.5$, 
with offsets up to 1~arcsec, the latter corresponds to a distance of  $\sim 7.5$~kpc at $z \sim 0.5$. They propose that  
89 of them may be AGNs (quasars and BL Lac) dislodged from their host galaxies' centers.  

\begin{figure}
\begin{center}
\includegraphics[width=12cm,angle=0]{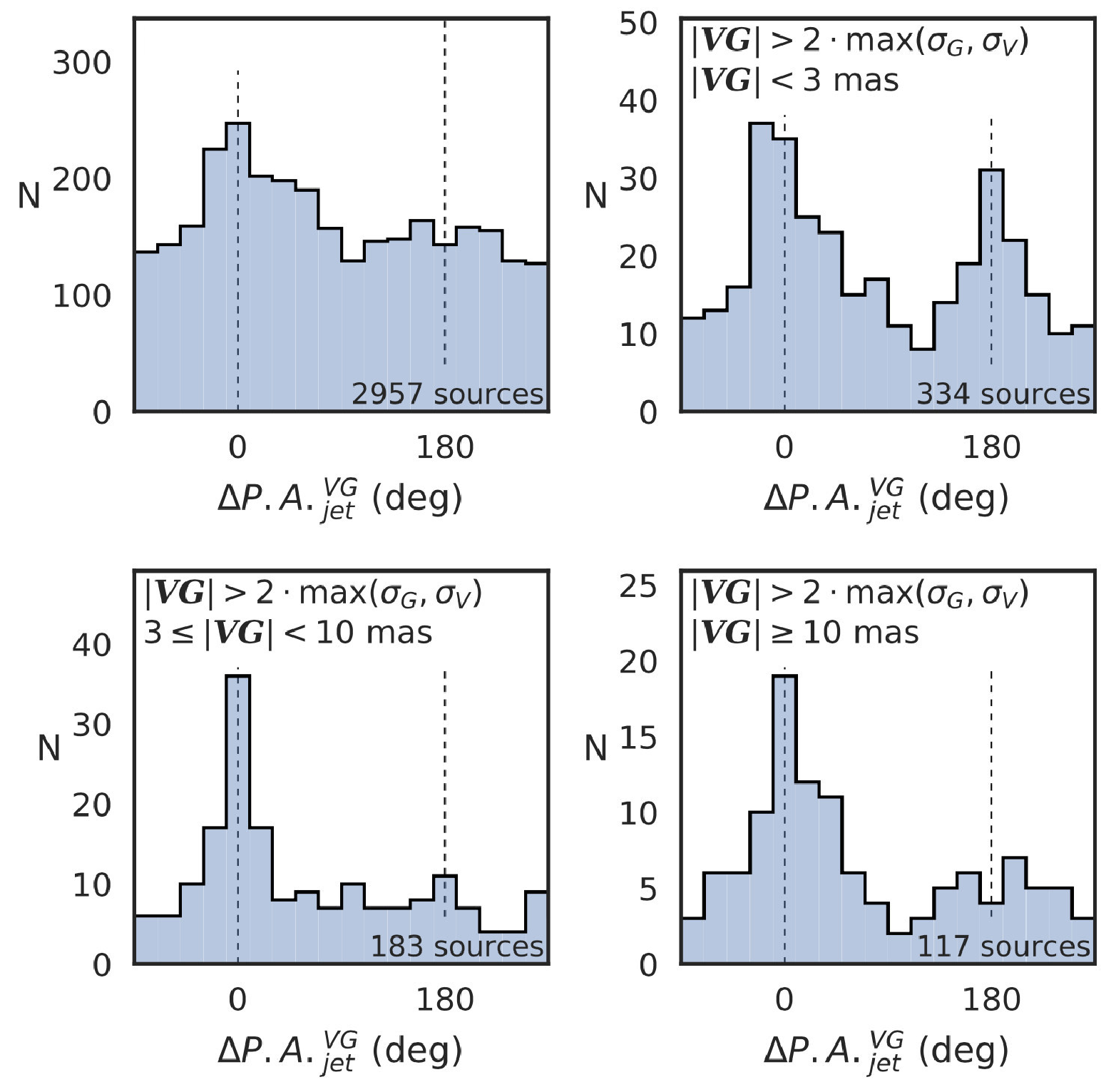}
\end{center}
\caption{Histograms of differences in position angle ($\Delta P.A.$) between the direction of the radio jet ({\it jet }) and the direction of the offset ($VG$) from radio to optical position of the source. All but the top-left panel plot only objects with the size of the offset $VG$ being at least two times larger than the errors of their VLBI ($\sigma_V$) and Gaia ($\sigma_G$) positions. Vertical dashed lines indicate aligned and counter-aligned jets. From \citet{kovalev17}, with permission.}\label{fig4}
\end{figure}

In a much larger study \citet{petrov17} used the whole 1.1~billion object dataset from Gaia DR1 and the VLBI absolute astrometry catalogue RFC 2016c to find 6055 secure matches with AGNs. \citet{kovalev17} used this large sample to check on offsets between radio and optical positions. In 2957 AGNs they were able to determine the direction of the radio jet. They find a significant prevalence of optical and radio offsets in directions along or opposite the one of the radio jet (Fig.~\ref{fig4}). This suggests that strong, extended optical jet structures are present in many AGN. Position offsets along the jet require strong, extended parsec-scale optical jets, while 
small ($< 1$~mas) offsets in direction opposite to the jet can be due to extended VLBI jet structure or a "core-shift" effect due to synchrotron opacity.

\section{Conclusions}

Gaia is living to its promise of revolutionising virtually any field of astronomy. In case of quasars two of the instruments on board are of particular interest: spectrophotometric BP$/$RP instrument and astrometric instrument. The former allows studies of SED variability for a complete sample of sources with $V<18$, and of general brightness variations and photometric redshift determination for all sources brighter than $V \sim 20$. The emphasis is on completeness, and on the ability to discover and characterise yet unknown quasars in this magnitude range. Both present a major advance to current lively research efforts to characterise quasar variability \citep[e.g.][]{peters16,marziani17,rumbaugh17}. An obvious importance of astrometry of quasars is to harmonise optical and radio reference frames. Some 135,000 quasars, which were already analysed within the first data release, already allowed to achieve an unprecedented accuracy of alignment of VLBI radio and optical Gaia reference frames. A small offset between optical and radio position, which is seen in some quasars with well determined directions of radio jets, points to presence of strong, extended optical jets on parsec scales. Spatial resolution of Gaia allows to discover large numbers of gravitationally lensed quasars \citep{lemon17}, especially when photometric and spectrophotometric measurements of Gaia are combined with mid-infrared selection, as enabled by the WISE mission \citep{agnello17}.

Gaia observations are ongoing, with the second public data release planned for April 2018. It will include a five-parameter astrometric solutions for all sources with acceptable formal standard errors ($> 10^9$ are anticipated), and positions (R.A., Dec.) for sources 
for which parallaxes and proper motions cannot be derived. Next, it will include $G$ and integrated BP and RP photometric  magnitudes for all sources. For sources brighter than $G_{RVS} = 12$ also median radial velocities will be published. Finally for stars brighter than $G = 17$ there will be estimates of the effective temperature and, where possible, line-of-sight extinction, based on the above photometric data. Of less interest to quasar studies, a list of photometric data for a  sample of variable stars and 
epoch astrometry for a pre-selected list of over 10,000 asteroids will be provided. 

There are further planned intermediate public data releases, with the final database to be published some 3 years after the end of observations, which may happen anytime between 2019 and 2023. By then a complete magnitude limited sample of quasars, many of them new, together with their variability, redshift and in many cases optical jets or multiple gravitationally lensed images are bound to be discovered.

\section*{Acknowledgments}
The author acknowledges financial support of the Slovenian Research Agency (research core funding No.\ P1-0188).
This research has been partially funded by the European Space Agency contract 4000111918.

\begin{table}
\caption{Properties of the Initial quasar catalogue with numbers of different sources and their expected typical precision, adapted from \citet{andrei14}. \medskip}
\label{table1}
\begin{center}
\begin{tabular}{lr}
\hline
Number of sources                              &1,248,372\\
Sources with magnitude                      & 1,246,512\\
Sources with redshift                           & 1,157,285 \\
Sources with $G < 20$                       &371,098\\
Sources with $G>21$                        &  690,507\\
Sources with $z < 1.0$                     & 250,405\\
Sources with $z > 2.0$                     & 383,487\\
Astrometry precision (arc~sec)            & 1\\
Magnitude precision                           & 0.5\\
Redshift precision                               &0.01\\
Average density (sources/deg$^2$)   &30.3\\
Average neighbour distance (arc~sec)&220\\\hline
\end{tabular}
\end{center}
\end{table}

\begin{thebibliography}{99}
\bibitem[Agnello (2017)]{agnello17} Agnello, A.\ 2017,  Quasar lenses and galactic streams: outlier selection and Gaia multiplet detection, \mnras, 471, 2013-2021
\bibitem[Andrei et al.\ (2014)]{andrei14} Andrei, A.H.; Ant\'{o}n, S.; Taris, F.; Bourda, G.; Souchay, J.; Bouquillon, J.; et al.\ 2014, The Gaia Initial Quasar Catalog, Proc. of the Journ\'{e}es 2013: "Syst\`{e}mes de r\'{e}f\'{e}rence spatio-temporels", Obs. de Paris, Nicole Capitaine (ed.), ISBN 978-2-901057-69-7, p. 84-87
\bibitem[Brown et al.\ (2016)]{brown16} Brown, A.G.A.; Vallenari, A.; Prusti, T.; de Bruijne, J.H.J.; Mignard, F.; Drimmel, R.; et al.\ 2016, Gaia Data Release 1. Summary of the astrometric, photometric, and survey properties, \aap, 595, A2, 23 pp.
\bibitem[Cui et al.\ (2012)]{cui12} Cui, Xiang-Qun; Zhao, Yong-Heng; Chu, Yao-Quan; Li, Guo-Ping; Li, Qi; Zhang, Li-Ping;  et al.\ 2012, The Large Sky Area Multi-Object Fiber Spectroscopic Telescope (LAMOST), Res. Astron. Astrophys., 12, 1197-1242
\bibitem[de Silva et al.\ (2015)]{deSilva15} De Silva, G. M.; Freeman, K. C.; Bland-Hawthorn, J.; Martell, S.; de Boer, E. Wylie; Asplund, M.; et  al.\ 2015, The GALAH survey: scientific motivation, \mnras, 449, 2604-2617
\bibitem[Gilmore et al.\ (2012)]{gilmore12} Gilmore, G.; Randich, S.; Asplund, M.; Binney, J.; Bonifacio, P.; Drew, J.; et al.\ 2012, The Gaia-ESO Public Spectroscopic Survey, Messenger, 147, 25-31
\bibitem[Kovalev et al.\ (2017)]{kovalev17} Kovalev, Y.Y.; Petrov, L.; Plavin, A.V. 2017, VLBI-Gaia offsets favor parsec-scale jet direction in active galactic nuclei, \aap, 598, L1, 4 pp.
\bibitem[Kunder et al.\ (2017)]{kunder17} Kunder, Andrea; Kordopatis, Georges; Steinmetz, Matthias; Zwitter, Toma\v{z}; McMillan, Paul J.; Casagrande, Luca;  et al.\ 2017, The Radial Velocity Experiment (RAVE): Fifth Data Release, \aj, 153, 75, 30 pp.
\bibitem[Lemon et al.\ (2017)]{lemon17} Lemon, Cameron A.; Auger, Matthew W.; McMahon, Richard G.; Koposov, Sergey E.\  2017, Gravitationally Lensed Quasars in Gaia: I. Resolving Small-Separation Lenses, \mnras, in print, arXiv:1709.08976, 10 pp.
\bibitem[Lindegren et al.\ (2016)]{lindegren16} Lindegren, L.; Lammers, U.; Bastian, U.; Hern\'{a}ndez, J.; Klioner, S.; Hobbs, D.; et al.\ 2016, Gaia Data Release 1. Astrometry: one billion positions, two million proper motions and parallaxes, \aap, 595, A4, 32 pp.
\bibitem[Makarov et al.\ (2017)]{makarov17} Makarov, Valeri V.; Frouard, Julien; Berghea, Ciprian T.; Rest, Armin; Chambers, Kenneth C.; Kaiser, Nicholas; et al.\ 2017, Astrometric Evidence for a Population of Dislodged AGNs, \apjl, 835, L30, 6 pp.
\bibitem[Martell et al.\ (2017)]{martell17} Martell, S. L.; Sharma, S.; Buder, S.; Duong, L.; Schlesinger, K. J.; Simpson, J.; et al.\ 2017, The GALAH survey: observational overview and Gaia DR1 companion, \mnras, 465, 3203
\bibitem[Marziani et al.\ (2017)]{marziani17} Marziani, P.; Bon, E.; Grieco, A.; Bon, N.; Dultzin, D.; Del Olmo, A.;  et al.\ 2017, Optical variability patterns of radio-quiet and radio-loud quasars, New Frontiers in Black Hole Astrophysics, Proc. IAU Symposium 324, 243-244, doi:	10.1017/S1743921316013065
\bibitem[Mignard et al.\ (2016)]{mignard16} Mignard, F.; Klioner, S.; Lindegren, L.; Bastian, U.; Bombrun, A.; Hern\'{a}ndez, J.; et al.\ 2016, Gaia Data Release 1. Reference frame and optical properties of ICRF sources, \aap, 595, A5, 16 pp.
\bibitem[Peters et al.\ (2016)]{peters16} Peters, Christina M.; Richards, Gordon T.; Myers, Adam D.; Strauss, Michael A.; Schmidt, Kasper B.; Ivezi\'{c}, \v{Z}eljko; et al.\ 2016, Quasar Classification Using Color and Variability, \apj, 811, 95, 29 pp.
\bibitem[Petrov \&\ Kovalev (2017)]{petrov17} Petrov, L., Kovalev, Y.Y. 2017, On significance of VLBI/Gaia position offsets, \mnras, 467, L71-L75
\bibitem[Proft et al.\ (2015)]{proft15} Proft, S., Wambsganss, J. 2015, Exploration of quasars with the Gaia mission, \aap, 574, A46, 11 pp. 
\bibitem[Prusti et al.\ (2016)]{prusti16} Prusti, T.; de Bruijne, J. H. J.; Brown, A. G. A.; Vallenari, A.; Babusiaux, C.; Bailer-Jones, C. A. L.;  et al.\ 2016, The Gaia mission, \aap, 595, A1, 36 pp.
\bibitem[Rumbaugh et al. (2017)]{rumbaugh17} Rumbaugh, Nick; Shen, Yue; Morganson, Eric; Liu, Xin; Banerji, Manda; McMahon, Richard G.;  et al.\ 2017, Extreme variability quasars from the Sloan Digital Sky Survey and the Dark Energy Survey, arXiv:1706.07875
\end{thebibliography}
\end{document}